\documentstyle[twoside,fleqn,espcrc2,epsf,epsbox,epsfig]{article}

\newcommand{\AmS}{{\protect\the\textfont2
A\kern-.1667em\lower.5ex\hbox{M}\kern-.125emS}}

\title{
Abelianization of QCD in the Maximally Abelian Gauge 
and the Nambu-'t~Hooft Picture for Color Confinement
}
\author{H.~Suganuma\address[TIT]{
\vspace{-0.05cm}
Faculty of Science, 
Tokyo Institute of Technology, 
Ohokayama 2-12-1, Meguro, Tokyo 152-8551, Japan}
and H.~Ichie\address[HB]{
\vspace{-0.05cm}
Humboldt Univ. zu Berlin, 
Institut f\"ur Physik, Invalidenstrasse 110, D-10115 Berlin, Germany}}
\begin{document}
\begin{abstract}
We study the Nambu-'t~Hooft picture for color confinement 
in terms of the abelianization of QCD and monopole condensation 
in the maximally abelian (MA) gauge. 
In the MA gauge in the Euclidean metric, the off-diagonal gluon amplitude is 
strongly suppressed, and then the off-diagonal gluon phase shows 
strong randomness, which leads to rapid reduction of the off-diagonal gluon correlation. 
In SU(2) and SU(3) lattice QCD in the MA gauge with the abelian Landau gauge, 
the Euclidean gluon propagator indicates 
a large effective mass of the off-diagonal gluon as $M_{\rm off} \simeq 1~{\rm GeV}$ 
in the intermediate distance as $0.2{\rm fm} \le r \le 0.8{\rm fm}$.
Due to the infrared inactiveness of off-diagonal gluons,  
infrared QCD is well abelianized like nonabelian Higgs theories in the MA gauge. 
We investigate the inter-monopole potential and the dual gluon field 
$B_\mu$ in the MA gauge, and find longitudinal magnetic screening 
with $m_B \simeq$ 0.5 GeV in the infrared region, 
which indicates the dual Higgs mechanism by monopole condensation.
We define the ``gluonic Higgs scalar field'' providing the MA projection, and 
find the correspondence between its hedgehog singularity and the monopole location in lattice QCD.
\vspace{-0.2cm}
\end{abstract} 

\maketitle

\section{Dual Superconductor Theory for Color Confinement and 
QCD in MA Gauge}

To understand color confinement is one of the most difficult problems remaining in particle 
physics \cite{CONF2000}. As the Regge trajectories and lattice QCD indicate, quark confinement is 
characterized by {\it one-dimensional squeezing} of the color-electric flux with  
the string tension $\sigma \simeq 1{\rm GeV/fm}$. 
On the confinement mechanism, Nambu 
proposed the {\it dual superconductor theory} in 1974, 
based on the electro-magnetic duality \cite{N74}. In this theory, there occurs the one-dimensional squeezing of the color-electric flux  
by the {\it dual Meissner effect} due to condensation of bosonic color-magnetic monopoles. 
But, there are {\it two large gaps} between QCD and the dual superconductor theory.
\begin{enumerate}
\item
The dual superconductor theory is based on the abelian gauge theory  
subject to the Maxwell-type equations, where electro-magnetic duality is 
manifest, while QCD is a nonabelian gauge theory.   
\vspace{-0.1cm}
\item
The dual superconductor theory requires color-magnetic monopole 
condensation as the key concept, while QCD itself does not have 
color-magnetic monopoles. 
\end{enumerate}
These gaps may be filled simultaneously by taking 
{\it maximally abelian (MA) gauge fixing,} which reduces QCD to an abelian gauge theory 
including color-magnetic monopoles \cite{tH81}.

In Euclidean QCD, the MA gauge is defined so as to minimize the ``total amount" of the 
off-diagonal gluon amplitude \cite{IS9900,S0098},
\begin{eqnarray}
\vspace{-0.3cm}
R_{\rm off} [A_\mu ( \cdot )] \equiv \int d^4x \ {\rm tr}
\left\{ 
[\hat D_\mu ,\vec H][\hat D_\mu ,\vec H]^\dagger 
\right\} 
\vspace{-0.3cm}
\end{eqnarray}
by the SU($N_c$) gauge transformation.
Since the ${\rm SU}(N_c)$ covariant derivative 
$\hat D_\mu \equiv \hat \partial_\mu+ieA_\mu $ obeys the 
adjoint gauge transformation, the local form of 
the MA gauge condition is easily derived as 
$
[\vec H, [\hat D_\mu , [\hat D_\mu , \vec H]]]=0. 
$

In the MA gauge, the nonabelian gauge symmetry is partially fixed as  
$
{\rm SU}(N_c)_{\rm local} \rightarrow  
{\rm U(1)}_{\rm local}^{N_c-1} 
\times {\rm Weyl}^{\rm global}_{N_c}, 
$
and QCD reduces into an abelian gauge theory, where 
the off-diagonal gluons behave as charged matter fields providing the 
color-electric current in terms of the residual abelian gauge symmetry. 
In the MA gauge, 
according to the reduction of the nonabelian gauge manifold, 
color-magnetic monopoles appear as the topological defects 
reflecting the nontrivial homotopy group
$
\Pi_2({\rm SU}(N_c)/{\rm U(1)}^{N_c-1})=\Pi_1({\rm U(1)}^{N_c-1})
={\bf Z}^{N_c-1}_\infty, 
$
as 't~Hooft \cite{tH81} pointed out with the similar logic on the 't~Hooft-Polyakov monopole.

Thus, in the MA gauge, {\it QCD reduces into an abelian gauge theory including color-magnetic monopoles,} 
and this situation is conjectured to provide a realistic dual-superconductor picture for the confinement mechanism based on QCD, 
which we call here the Nambu-'t~Hooft picture.

\section{Strong Randomness of Off-diagonal Gluon Phases in the MA Gauge 
and Its Physical Implications}

To find out essence of the MA gauge on abelianization,  
we study the off-diagonal gluon field 
$
A_\mu^\pm(x) \equiv \frac1{\sqrt{2}}(A_\mu^1 \pm i A_\mu^2) 
=e^{\pm i \chi_\mu(x)}|A_\mu^\pm(x)|
$
in the MA gauge in SU(2) lattice QCD, and 
find the two remarkable features as follows \cite{IS9900,S0098}. 
\begin{enumerate}
\item
The off-diagonal gluon amplitude $|A_\mu^{\pm}(x)|$ 
is strongly suppressed by SU($N_c$) gauge transformation in the MA gauge.
\vspace{-0.1cm}
\item
The off-diagonal gluon phase $\chi_\mu(x)$ tends 
to be random, because $\chi_\mu(x)$ is not 
constrained by MA gauge fixing at all, 
and only the constraint from the QCD action is weak 
due to a small accompanying factor $|A_\mu^\pm|$.
\vspace{-0.1cm}
\end{enumerate}
Owing to the strong randomness of the off-diagonal gluon phase, 
we expect strong cancellation of the off-diagonal gluon contribution in the MA gauge,  
especially for the infrared quantities.

For instance, within the random-variable approximation 
for the off-diagonal gluon phase $\chi_\mu(s)$ in the MA gauge, 
we analytically derive 
{\it perfect abelian dominance of the string tension,} 
$\sigma_{\rm SU(2)}=\sigma_{\rm Abel}$, 
via the {\it perimeter law} of 
the off-diagonal gluon contribution 
in the Wilson loop~as  
\begin{eqnarray}
W_C^{\rm off} &\equiv& {\langle W_C[A_\mu^a]\rangle}/
{\langle W_C[A_\mu^\pm \equiv 0, A_\mu^3]\rangle_{\rm MA}} \nonumber \\
&\simeq& \exp\{-L a^2 \langle |eA_\mu^\pm|^2 \rangle_{\rm MA}/4\}, 
\end{eqnarray}
where 
$
\langle W_C[A_\mu^a]\rangle 
$
denotes the SU(2) Wilson loop,  
and 
$
\langle W_C[A_\mu^\pm \equiv 0, A_\mu^3]\rangle_{\rm MA}
$
the {\it abelian Wilson loop} in the MA gauge.
We check  this relation between the {\it macroscopic} quantity $W_C^{\rm off}$ 
and the {\it microscopic} quantity 
$\langle |eA_\mu^\pm|^2 \rangle_{\rm MA}$ in lattice QCD \cite{IS9900,S0098}. 

\begin{figure}[htb]
\vspace{-0.5cm}
\begin{center}
\epsfig{figure=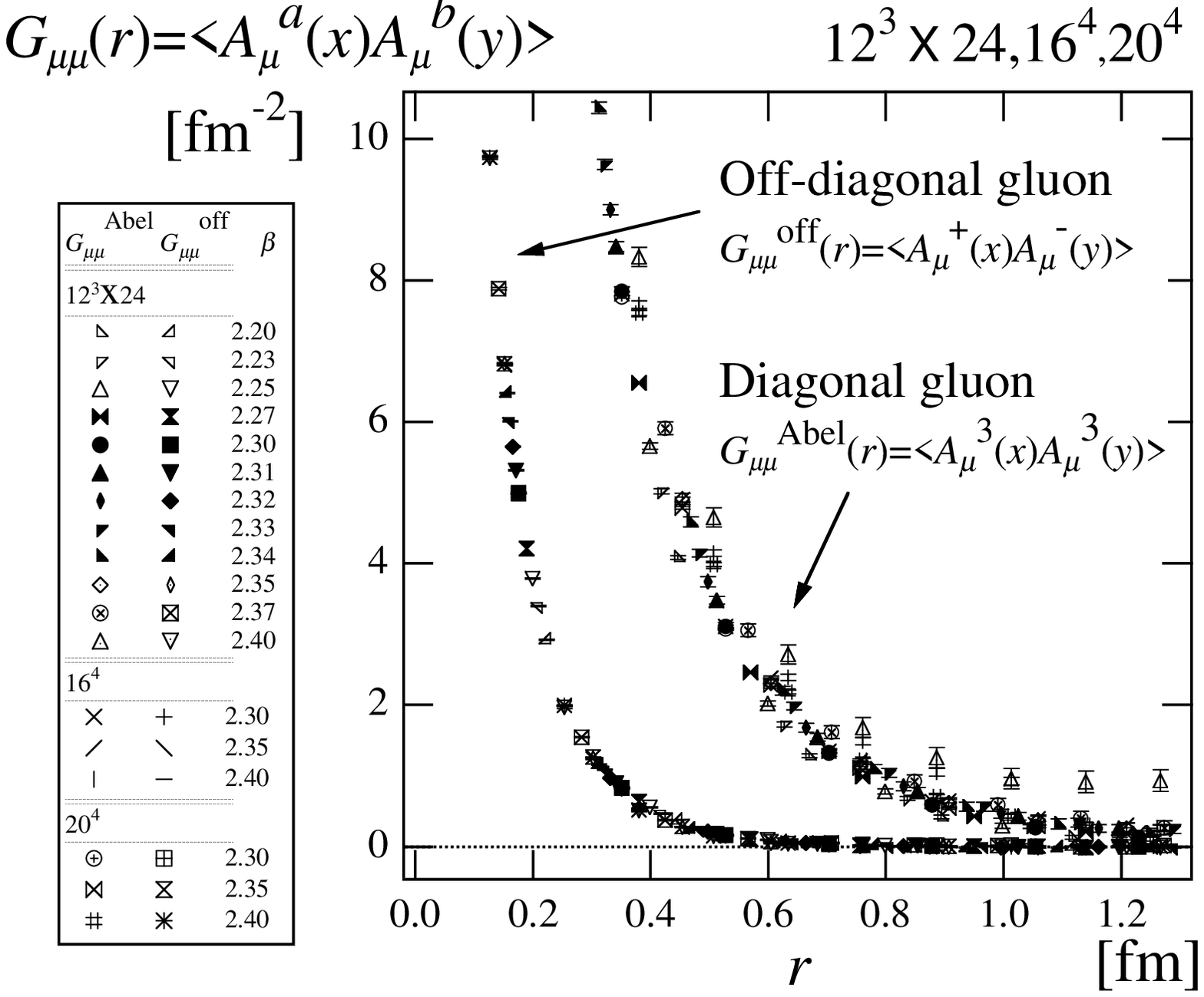,height=4.81cm} 
\epsfig{figure=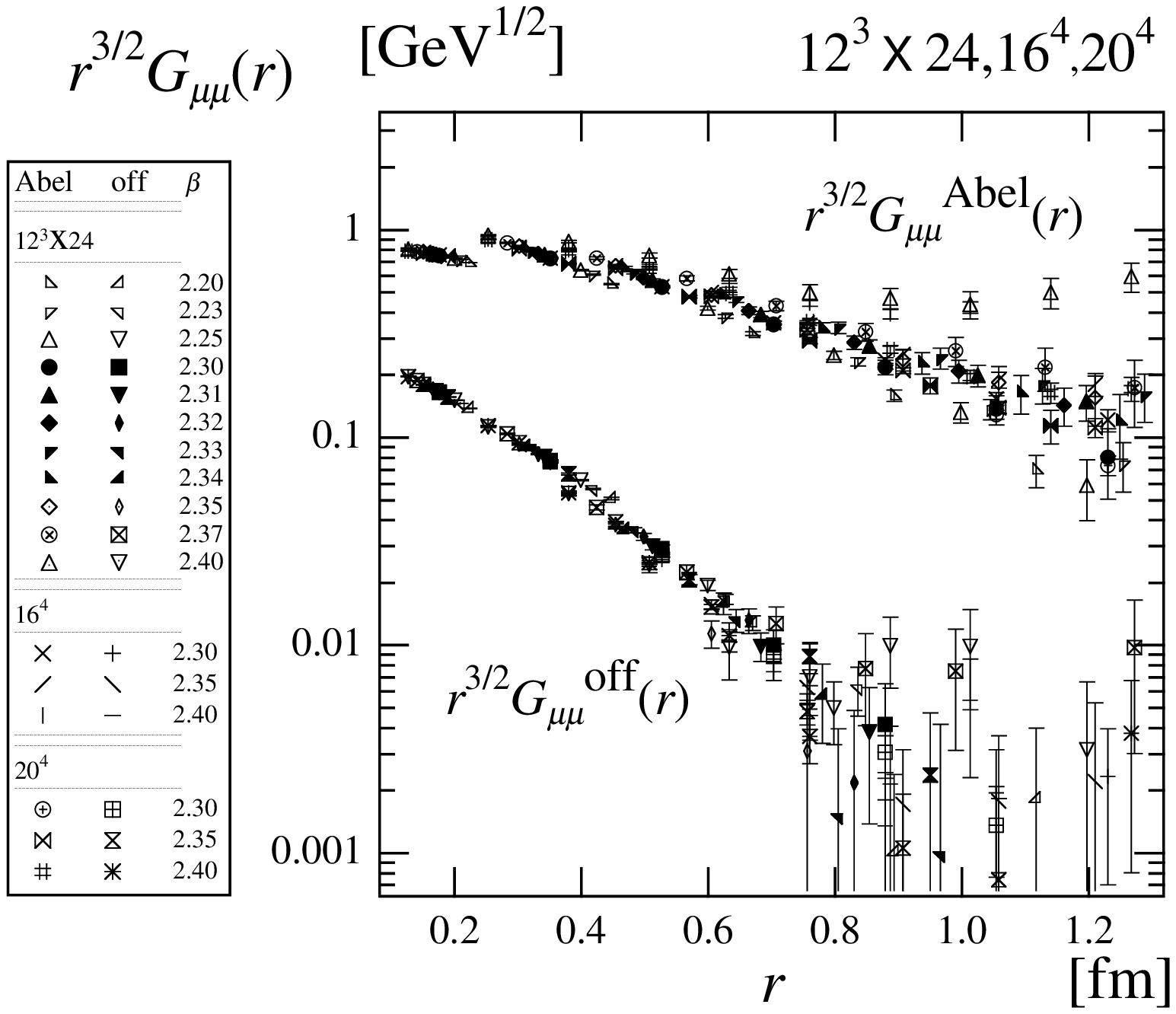,height=4.81cm}
\vspace{-1.0cm}
\caption{(a) The scalar-type gluon propagator 
$G_{\mu \mu }^a(r)$ v.s. 4-dim.~distance $r$ 
in the MA gauge in SU(2) lattice QCD with  
$2.2 \le \beta \le 2.4$, $12^3 \times 24$, $16^4$, $20^4$. 
(b) The logarithmic plot of $r^{3/2} G_{\mu \mu}^a(r)$. 
}
\end{center} 
\vspace{-0.75cm}
\end{figure}

As another physical implication, 
{\it strong randomness of off-diagonal gluon phases} 
in the MA gauge leads to   
{\it rapid reduction of off-diagonal gluon correlations}   
\cite{S0098} as 
\begin{eqnarray}
\langle A_\mu^+(x) A_\nu^-(y) 
\rangle_{\rm MA} \!\!\!\!\!&=&\!\!\!\!\! 
\langle |A_\mu^+(x)A_\nu^-(y)| e^{i\{\chi_\mu(x)-\chi_\nu(y)\}} 
\rangle \nonumber \\ 
\!\!\!\!\!&\sim&\!\!\!\!\! \langle |A_\mu^\pm(x)|^2 \rangle_{\rm MA} 
\delta_{\mu\nu}\delta^4(x-y), 
\vspace{-0.1cm}
\end{eqnarray}
which physically indicates a large effective mass of the off-diagonal gluon in the MA gauge.

\section{Large Effective-Mass Generation of Off-diagonal Gluons in MA Gauge : 
Essence of Infrared Abelianization of QCD}

We study the Euclidean gluon propagator in the MA gauge 
\cite{S0098,AS99}
using SU($N_c$) lattice QCD with $N_c=2,3$.
On the residual abelian gauge symmetry, 
we take the abelian Landau gauge, where 
the gluon configuration becomes maximally continuous   
under the MA gauge constraint. 

\begin{figure}[htb]
\vspace{-0.5cm}
\begin{center}
\epsfig{figure=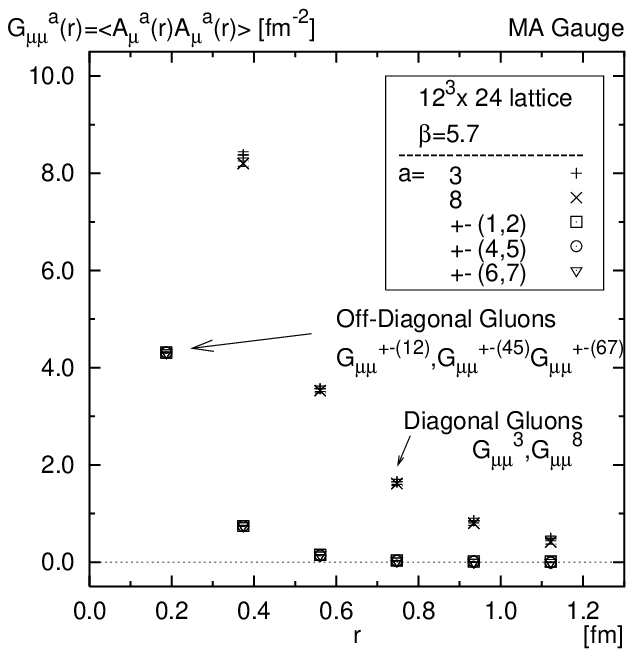,height=4.85cm}
\epsfig{figure=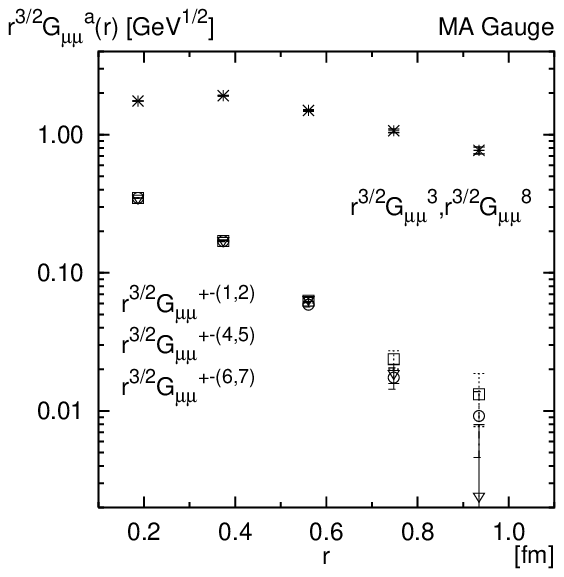,height=4.85cm}
\vspace{-1.0cm}
\caption{
(a) The scalar-type gluon propagator 
$G_{\mu \mu }^a(r)$ v.s. the four-dimensional 
distance $r$ in the MA gauge 
with ${\rm U(1)}_3 \times {\rm U(1)}_8$ Landau gauge fixing 
in SU(3) lattice QCD with
$\beta=5.7$ and $12^3 \times 24$.
(b) The logarithmic plot of $r^{3/2} G_{\mu \mu}^a(r)$. 
}
\end{center}
\vspace{-0.85cm}
\end{figure}

We show the scalar-type gluon propagators $G_{\mu \mu}^{a}(r) 
\equiv \langle A_\mu^{a}(x)A_\mu^{a}(y)\rangle$ ($a=1,2,.., N_c^2-1$) 
as the function of the four-dimensional Euclidean 
distance $r \equiv \sqrt{(x_\mu- y_\mu)^2}$ in Figs.1 and 2.
Both in SU(2) and SU(3) QCD, 
we find {\it infrared abelian dominance for the gluon propagator 
in the MA gauge}:
while the diagonal gluon, $A_\mu^3(x)$ or $A_\mu^8(x)$, propagates over 
the long distance, off-diagonal gluons propagate only within a short distance.

Next, we evaluate the effective off-diagonal gluon mass in the MA gauge 
from the slope analysis of the lattice QCD data of 
$\ln\{r^{3/2}G_{\mu\mu}(r)\}$, since 
the four-dimensional Euclidean propagator of the 
massive vector boson with the mass $M$ takes a 
Yukawa-type asymptotic form as 
\begin{eqnarray}
\vspace{-0.2cm}
G_{\mu\mu}(r) \simeq \frac3{4\pi^2} \frac{M}{r} K_1(Mr)
\simeq \frac{3M^{1/2}}{2(2\pi)^{3/2}}\frac{e^{-Mr}}{r^{3/2}}.
\vspace{-0.2cm}
\end{eqnarray}
As a remarkable fact, the off-diagonal gluon behaves as a massive vector field with 
a large constant effective mass $M_{\rm off}$ in the intermediate distance as $0.2 {\rm fm} \le r \le 0.8 {\rm fm}$ in the MA gauge:
\begin{enumerate}
\item
$M_{\rm off} \simeq 1.2~{\rm GeV}$ 
in SU(2) lattice QCD 
with $2.2 \le \beta \le 2.4$ 
and $12^3 \times 24$, $16^4$, $20^4$ (Fig.1).
\item
$M_{\rm off} \simeq 1.1~{\rm GeV}$ 
in SU(3) lattice QCD with $\beta=5.7$ ($a \simeq $ 0.19fm) 
and $12^3 \times 24$ (Fig.2). 
\end{enumerate}
Thus, both in SU(2) and SU(3) QCD, 
the {\it off-diagonal gluon seems to acquire   
a large effective mass 
$M_{\rm off} \simeq 1 {\rm GeV}$ in the MA gauge}, which would be  
{\it essence of infrared abelian dominance} 
\cite{S0098,AS99}.

\section{Inter Monopole Potential, Longitudinal Magnetic Screening,   
Infrared Monopole Condensation and Monopole Structure}

Using SU(2) lattice QCD, we study 
the {\it inter-monopole potential} and 
the {\it dual gluon propagator} 
in the monopole part in the MA gauge,  
and show {\it longitudinal magnetic screening} in the infrared region, 
as a direct evidence of 
the {\it dual Higgs mechanism by monopole condensation.} 
The dual gluon mass is estimated as $m_B \simeq$ 0.5 GeV  
\cite{S0098}.
Then, lattice QCD in the MA gauge exhibits 
{\it infrared abelian dominance} and 
{\it infrared monopole condensation}, 
which lead to 
the dual Ginzburg-Landau (DGL) theory \cite{SST95} for infrared QCD.

Using SU(2) lattice QCD, we find the structure of the monopole in the MA gauge 
like the 't~Hooft-Polyakov monopole: a large off-diagonal gluon amplitude 
around its center, as shown in Fig.3.  
At a large scale, off-diagonal gluons 
inside monopoles become invisible, 
and monopoles can be regarded as point-like Dirac monopoles 
\cite{IS9900,S0098}. 

\begin{figure}[htb]
\vspace{-1.05cm}
\begin{center}
\epsfig{figure=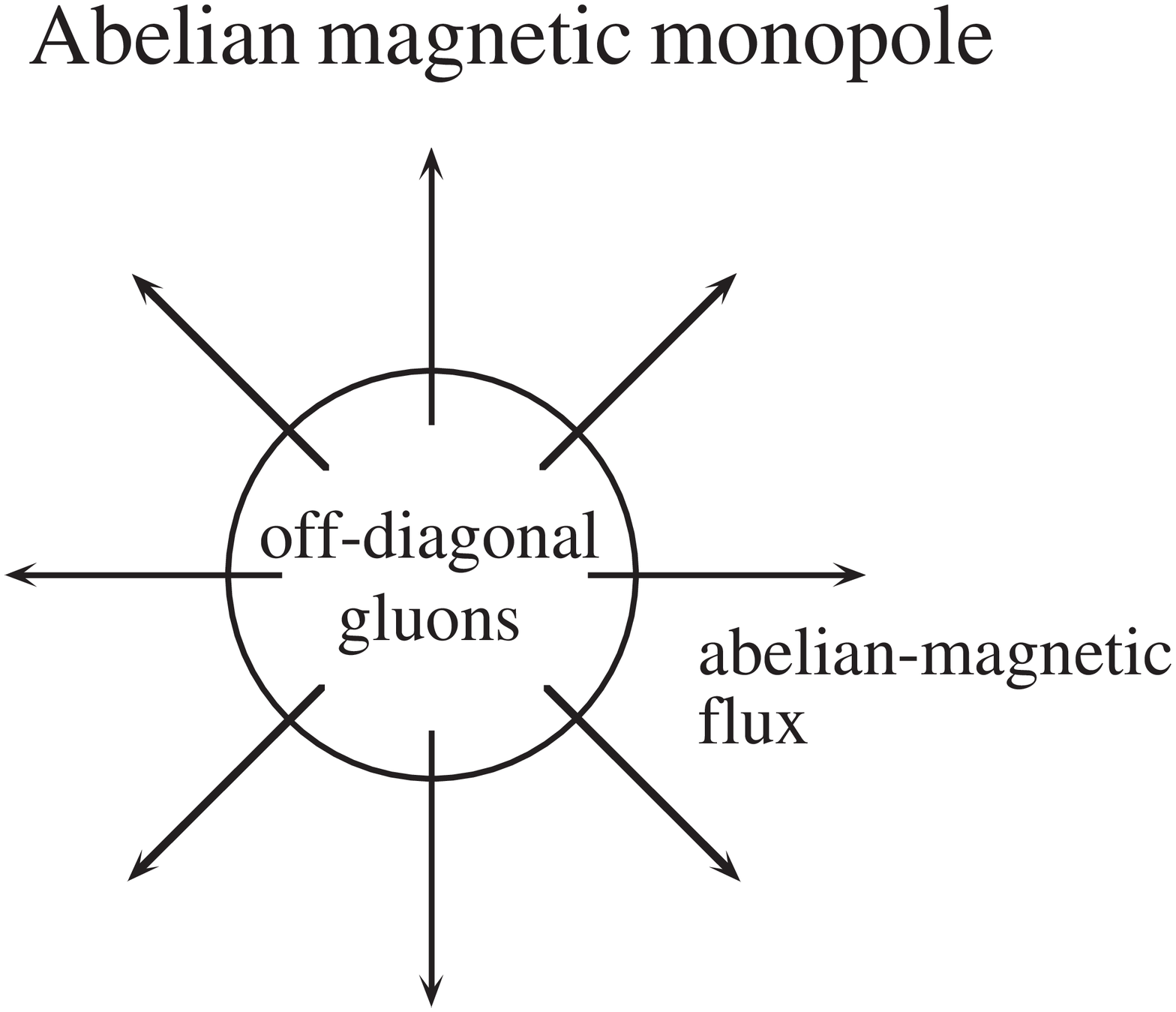,height=0.5\columnwidth}
\vspace{-1.0cm}
\caption{
The schematic figure of the structure of the monopole in the MA gauge.
The off-diagonal gluon amplitude is large around its center as well as the abelian magnetic field. 
}
\end{center}
\vspace{-0.4cm}
\end{figure}

\section{Monopole in MA Gauge and 
Hedgehog Singularity of the Gluonic Higgs Field}

To clarify the similarity between QCD in the MA gauge and the NAH theory, 
we introduce the ``gluonic Higgs scalar field'' 
$\vec \phi(x) \equiv \Omega(x) \vec H \Omega^\dagger(x)$ 
with $\Omega(x) \in {\rm SU}(N_c)$ 
so as to minimize 
\begin{eqnarray}
\vspace{-0.55cm}
R[\vec \phi(\cdot)] \equiv \int d^4x \ {\rm tr} 
\left\{[\hat D_\mu, \vec \phi(x)][\hat D_\mu, \vec \phi(x)]^\dagger \right\}
\vspace{-1.55cm}
\end{eqnarray}
for arbitrary given Euclidean gluon field $\{A_\mu(x)\}$. 
The gluonic Higgs scalar $\vec \phi(x)$ physically corresponds to 
a ``color-direction'' of the nonabelian gauge connection $\hat D_\mu$ 
averaged over $\mu$ at each $x$. 

Similar to $\hat D_\mu$, 
$\vec \phi(x)$ obeys the adjoint gauge transformation, 
and $\vec \phi(x)$ is diagonalized in the MA gauge. 
Therefore, $\vec \phi(x)$ behaves as the Higgs scalar in the NAH theory, and 
{\it the hedgehog singularity of $\vec \phi(x)$ provides the monopole in the MA gauge.} 
This correlation is observed in lattice QCD, 
when the gluon field is continuous as in the SU(2) Landau gauge, as shown in Fig.4 \cite{IS9900,S0098}.

Through the projection along $\vec \phi(x)$, 
one can extract the abelian U(1)$^{N_c-1}$ sub-gauge-manifold 
close to the original SU($N_c$) gauge manifold. 
This projection is manifestly gauge invariant and is equivalent to the ordinary 
MA projection. Hence, {\it infrared relevance of the gluon mode along 
the color-direction $\vec \phi(x)$} is observed \cite{IS9900,S0098}.

\vspace{0.2cm}

\noindent
{\bf Acknowledgement } 
H.S. would like to thank Prof. G.~'t~Hooft and Prof. Yoichiro~Nambu for their valuable comments and discussions.

\begin{figure}[htb]
\vspace{-0.5cm}
\begin{center}
\epsfig{figure=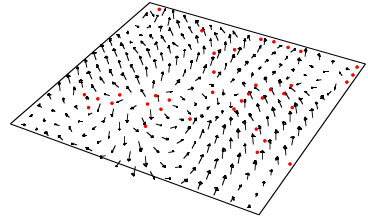,height=0.38\columnwidth}
\vspace{-1.0cm}
\caption{
The gluonic Higgs scalar field $\phi(x)=\phi^a(x)\frac{\tau^a}{2}$ 
in the SU(2) Landau gauge in SU(2) lattice QCD with $\beta=2.4$ and $16^4$.
The arrow denotes $(\phi^1(x),\phi^2(x),\phi^3(x))$.
The monopoles (dots) in the MA gauge appear at the hedgehog singularities of 
the gluonic Higgs scalar $\phi(x)$. 
}
\end{center}
\vspace{-0.9cm}
\end{figure}

\noindent
{\bf DISCUSSIONS}

\vspace{0.2cm}

As 't~Hooft pointed out, 
if the dual superconductor picture is true, the off-diagonal color-charges are to be confined, 
and the off-diagonal gluon correlation should be cut in the infrared limit.
This tendency is observed in Figs.1~and~2 as the rapid vanishing of off-diagonal gluon correlators in $r \ge 1 {\rm fm}$.
This infrared screening  would occur due to the off-diagonal color-charge confinement and the off-diagonal gluon pair creation.

So, the effective off-diagonal gluon mass $M_{\rm off} \simeq 1{\rm GeV}$ 
in the MA gauge is to be interpreted as the ``constituent mass'' 
in the intermediate distance, $0.2{\rm fm} \le r \le 0.8 {\rm fm}$, as 't~Hooft suggested. 

As an interesting possibility, we conjecture 
the mutual relation among the localization of the off-diagonal gluon correlation, abelian monopole condensation and 
the off-diagonal color-charge confinement in the MA gauge as shown in Fig.5.

\begin{figure}
\vspace{-1.75cm}
\epsfig{figure=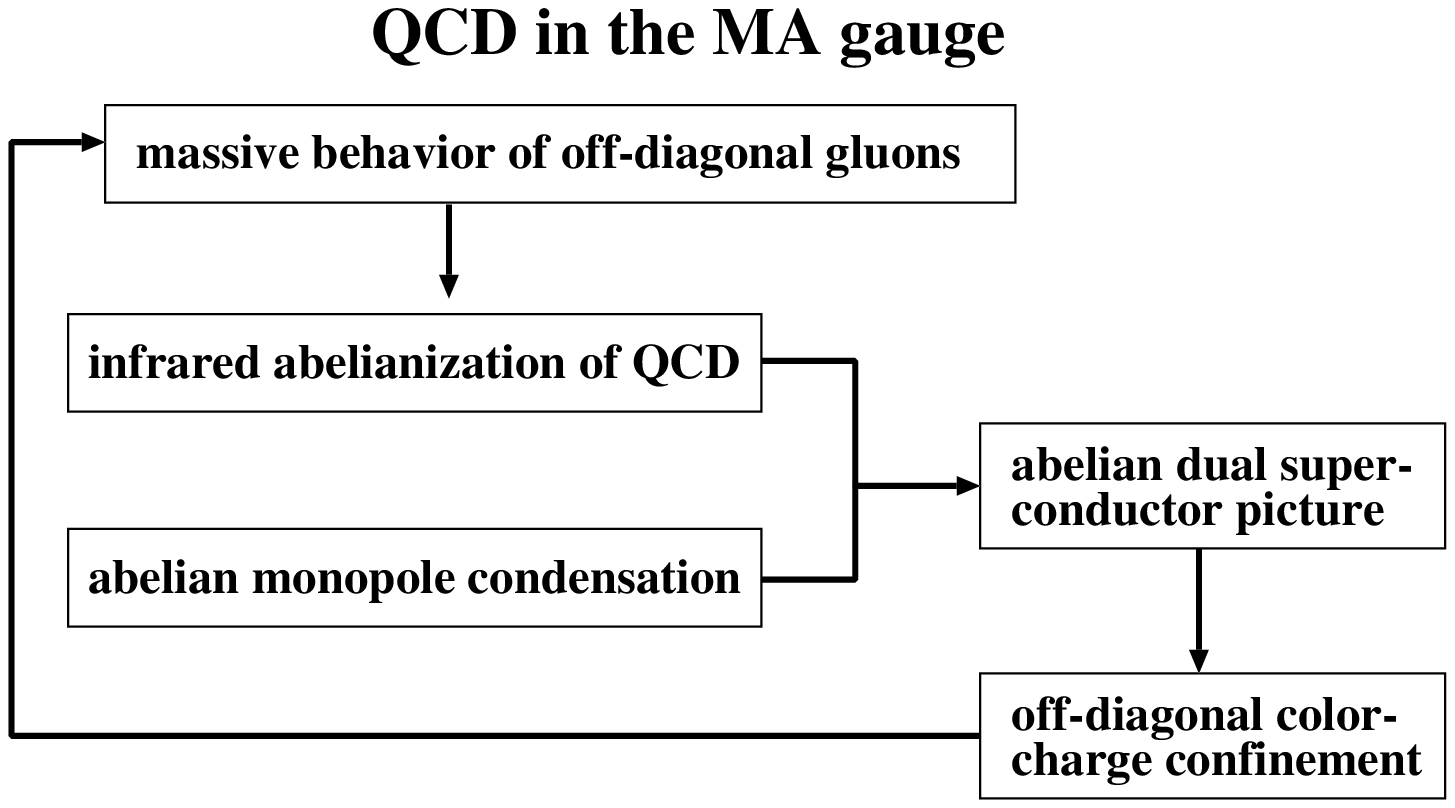, width=8.5cm}
\vspace{-1.3cm}
\caption{
The conjecture on the relation between abelianization of QCD and the off-diagonal color-charge confinement in the MA gauge.
}
\vspace{-0.0cm}
\end{figure}
\end{document}